# On the scaling of electron dechanneling length in bent crystals

V.M. Biryukov

Institute for High Energy Physics, 142281 Protvino, Russia

**Abstract**

We suggest a way to compare the experiments and simulations of electron dechanneling length in bent crystals performed at different energies and different radii. As an example, we compare the SLAC experiment at 3-14 GeV with earlier published predictions of two different Monte Carlo codes for 855 MeV and 50 GeV, and with MAMI experiment at 855 MeV.

## 1. Introduction

Following the idea of Tsyganov [1] to apply bent crystals for beam deflection, the interaction of particles with bent crystals has been studied in detail at many accelerator centers [2-16]. Pioneering experiment at Dubna [2] has demonstrated the effect of deflection with protons of a few GeV energy. Later the energy range of bent crystal experiments was extended down to 3 MeV and up into high energy frontier [17-25]. A variety of positive particles was used in channeling experiments in bent crystals, from positrons to heavy ions in a wide energy range [26-32].

First applications of bent-crystal channeling technique for beam delivery at accelerators have started in the 1980s, including beam extraction from accelerator rings, beam splitting and deflection in external beam lines [33-39]. Crystal fields are equivalent to order of 1000 Tesla. The largest angle of deflection that was used in practical applications at accelerators amounted to 150 mrad [36] (i.e. 9$^o$). At IHEP, a 150-mrad bent crystal was used to create a test area, delivering 70-GeV protons with intensity $10^6$ p/s to particle experiments. This 100-mm long crystal was equivalent to about 32 Tesla-meter.

Six locations on IHEP 70-GeV ring are equipped by crystal extraction systems. Channeling efficiency of about 85% is obtained at intensities in the order of $10^{12}$ proton/s [40,41]. The success of crystal applications has stimulated many proposals for crystal-assisted beam steering, in particular at the LHC energy [42-46]. Simulations predict high efficiency for LHC beam extraction [47-48] and collimation [49-50] assisted by bent crystals, while recent experiments demonstrated that bent crystals can be very efficient at LHC energies [25].

Channeling of positive projectiles in bent crystals is reasonably well understood theoretically. Analytical models and Monte Carlo simulations are able to predict the results of experiments [7,13,41,51-54]. More recent development is the experiments with negative particles channeled in bent crystals [55-59]. Several experiments were made with electrons of sub-GeV and multi-GeV energies, measuring the dechanneling lengths of negative projectiles in bent crystals at different energies *E* and bending radii *R*. The physics of electron channeling in bent crystals still needs a better understanding. There is not enough clarity for dependence of electron dechanneling length on particle energy and on the crystal bending radius. With established theoretical models we could design many applications for negative particles as well.

Can the studies of electron dechanneling in bent crystals, in experiments or Monte Carlo simulations, be extrapolated onto different energies and bending radii? How far in these extrapolations we could go? How strong is the change in the behavior of an electron channeled in a bent crystal if we increase (or reduce) the electron energy by an order of magnitude? What is the accuracy of such extrapolations? We look for the answers to these questions basing on the electron dechanneling data available from the experiments at SLAC [59] and MAMI [58], and from simulations with two different Monte Carlo codes.

## 2. The assumptions on dechanneling length behavior in bent crystal

A particle can be channeled if it enters the crystal lattice nearly parallel (within the Lindhard critical angle $\pm\theta_c$) to a major crystallographic direction, plane or axis. The critical angle $\theta_c$ equals

$$\theta_c = (2U_0/pv)^{0.5}, \tag{1}$$

where $U_0$ is the depth of the transverse potential well that confines the channeled particle with momentum $p$ and velocity $v$. In the following we replace $pv$ with energy $E$ as we discuss here only ultrarelativistic particles.

Once trapped, the channeled particles undergo multiple and single scatterings on the crystal constituents, nuclei and electrons. This scattering causes a gradual loss, *dechanneling*, of particles from the channeled states to random states. This loss is usually an exponential decay characterized by a dechanneling length $L_D$.

Baier et al. [60] estimated the electron dechanneling length $L_D$ in crystals comparing the angle of multiple scattering $\theta_{ms}$ in crystal over length $L$ with the critical angle of channeling $\theta_c$. For negative projectiles, they assumed the angle of multiple scattering in crystal be the same as in amorphous body. They take the *rms* angle $\theta_{ms}$ of multiple scattering over distance $L$ as

$$\theta_{ms} = E_s/E \cdot (L/X)^{0.5} \tag{2}$$

where constant $X$ is radiation length, and $E_s \approx 13.6$ MeV. The $L_D$ value was defined by Baier et al. as the length over which the multiple scattering angle equals the critical angle, $\theta_{ms} \approx \theta_c$:

$$E_s/E \cdot (L/X)^{0.5} \approx (2U_0/E)^{0.5} \tag{3}$$

Then

$$L_D \approx (2U_0 E/E_s^2) \cdot X \tag{4}$$

Baier et al. use $E_S$ in the definition of Rossi-Greisen

$$E_S = m_e c^2 \cdot \sqrt{2\pi/\alpha} \tag{5}$$

for the projection angle. Here $m_e$ is the electron mass at rest, $\alpha \approx 1/137$. Their ultimate formula is

$$L_D \approx (\alpha U_0 E/\pi m_e^2 c^4) \cdot X \tag{6}$$

This model predicts a linear dependence of dechanneling length $L_D$ on energy. It has been widely used over recent four decades for comparison with the experimental results.

In a bent crystal, a centrifugal force appears in the equation of motion of a particle in bent crystal planes [61]

$$E\,(d^2x/dz^2) + U'(x) + E/R(z) = 0$$

Here $U'(x) = dU(x)/dx$ is the gradient of the interplanar potential $U(x)$ and $x$ is the transverse coordinate. The potential well $U(x)$ where channeled particles are trapped in a straight crystal transforms in a bent crystal into an effective potential $U_{eff}(x)$ with centrifugal term $Ex/R$.

$$U_{eff}(x) = U(x) + Ex/R$$

In particular, the depth of the potential well now depends on the bending radius $R$. In the equations, the radius appears solely through the ratio $E/R$. As the radius $R$ reduces gradually and approaches the Tsyganov critical radius $R_c$, the depth of the potential well reduces as well, and disappears when $R \leq R_c$.

As a result, the dechanneling length in a bent crystal should depend on the bending radius. Based on the above considerations, we assume that the electron dechanneling length $L_D$ in bent crystals takes the form

$$L_D(E,R) = L_{D0}(E) \cdot f(E/R), \qquad (7)$$

where $L_{D0}(E)$ is the dechanneling length in a straight crystal, and $f(E/R)$ is *some* function which depends on $R$ solely through the ratio $E/R$. We don't need to specify what this $f(E/R)$ function is. For us it will be enough to rely solely on the assumption that if $E_2/R_2 = E_1/R_1$, then $f(E_2/R_2) = f(E_1/R_1)$. This assumption (that doesn't specify the $f(E/R)$ function) should make the comparisons more reliable and independent of the hypotheses of what the function $f(E/R)$ may be.

We also assume for the moment that $L_{D0}(E)$ is a linear function of energy within the range that involves the compared datasets. We don't need to know the exact proportionality coefficient between $L_{D0}$ and $E$. These two assumptions are sufficient to make possible the comparisons of different datasets obtained in different energy ranges with different bending radii.

Then, the dechanneling lengths $L_D(E,R)$ for the pairs $(E,R)$ with equal ratios $E_2/R_2 = E_1/R_1$ must correspond to each other through the ratio

$$L_D(E_2,R_2) = L_D(E_1,R_1) \cdot (E_2/E_1) = L_D(E_1,R_1) \cdot (R_2/R_1) \qquad (8)$$

Table 1 shows the datasets obtained in the experiments at SLAC with electrons of 3.35 to 14 GeV and MAMI with electrons of 855 MeV, and in simulations with two different Monte Carlo codes which made predictions for 50 GeV and 855 MeV electrons. We shall use these datasets to draw some conclusions below.

|  |  | $E$ (GeV) | $R$ (cm) | $E/R$ (GeV/cm) | $L_D$ (µm) | $L_D$ dyn. (µm) |
|---|---|---|---|---|---|---|
| SLAC | 2016 | 3.35 | 15 | 0.22 | 55.4 | - |
| SLAC | 2016 | 4.2 | 15 | 0.28 | 45.2 | - |
| SLAC | 2016 | 6.3 | 15 | 0.42 | 65.3 | - |
| SLAC | 2016 | 10.5 | 15 | 0.7 | 57.5 | - |
| SLAC | 2016 | 14 | 15 | 0.93 | 55.8 | - |
| Biryukov | 2007 | 50 | 41.67 | 1.2 | 110 | - |
| MAMI | 2014 | 0.855 | 3.35 | 0.26 | 19.2 | - |
| Mazzolari et al. | 2014 | 0.855 | 0.8 | 1.07 | 6.9 | 4.9 |
| Mazzolari et al. | 2014 | 0.855 | 1.6 | 0.53 | 13.6 | 10.2 |
| Mazzolari et al. | 2014 | 0.855 | 3.3 | 0.26 | 19.5 | 13.6 |
| Mazzolari et al. | 2014 | 0.855 | 6.7 | 0.13 | 25.5 | 15.4 |
| Mazzolari et al. | 2014 | 0.855 | 13.4 | 0.06 | 35.5 | 16.2 |

**Table 1** Datasets obtained in the experiments at SLAC with electrons of 3.35 to 14 GeV and MAMI with electrons of 855 MeV, and in simulations with two different Monte Carlo codes which made predictions for 50 GeV and 855 MeV electrons.

## 3. Comparisons of dechanneling length from different experiments and simulations

Let us consider an example of the Monte Carlo simulation [62] made with electrons at $E_1 = 50$ GeV for Si (111) crystal bent with radius $R_1 = 41.7$ cm and $E_1/R_1 = 1.2$ GeV/cm. The simulation has found $L_D = 110$ µm. This simulation used the code CATCH [63,64] which was frequently applied for simulations of bent crystal channeling of protons and ions [7-9,11,13,22,28,32,38-41,49-53]. CATCH has no free parameters. The code tracks a charged particle through the distorted-crystal lattice with the use of continuous-potential approximation and the non-diffusion approach to the processes of multiple and single scattering on the crystal consituents, electrons and nuclei, as well as on the crystal dislocations if necessary [65].

In this simulation, an electron was allowed to dechannel and rechannel, i.e. to return from a random (overbarrier) state back to a channeled (underbarrier) state any number of times. This is done in order to comply with the particle behavior in experiments. The overall exit angular distribution in simulation was found exponential, $N(\theta) \sim \exp(-R_1\theta/L_D)$. It was fitted to obtain a dechanneling length $L_D$ that characterized the exponential decay.

The Si (111) crystal tested in the experiments [59] at SLAC was bent with radius $R_2 = 15$ cm. This means that an electron with energy $E_2 = E_1 R_2/R_1$ would have in that crystal the same ratio $E/R$ as in the above simulation. Then, in accordance with Eq. (2) we expect that dechanneling length $L_D$ equals $110 \cdot (15/41.7) = 39.6$ µm at energy $E_2 = 50 \cdot (15/41.7) = 18$ GeV and $R_2 = 15$ cm, if we extrapolate that simulation result on the SLAC crystal.

We show this prediction in Table 2 and in Figure 1 together with SLAC experimental data.

|  |  | $E$ (GeV) | $R$ (cm) | $E/R$ (GeV/cm) | $L_D$ (μm) | $L_D$ dyn. (μm) |
|---|---|---|---|---|---|---|
| Biryukov | 2007 | 18 | 15 | 1.2 | 39.6 | - |
| MAMI | 2014 | 3.83 | 15 | 0.26 | 86 | - |
| Mazzolari et al. | 2014 | 16.03 | 15 | 1.07 | 129.4 | 91.9 |
| Mazzolari et al. | 2014 | 8.02 | 15 | 0.53 | 127.5 | 95.6 |
| Mazzolari et al. | 2014 | 3.89 | 15 | 0.26 | 88.6 | 61.8 |
| Mazzolari et al. | 2014 | 1.91 | 15 | 0.13 | 57.1 | 34.5 |
| Mazzolari et al. | 2014 | 0.96 | 15 | 0.06 | 39.7 | 18.1 |

**Table 2** Datasets of MAMI and Monte Carlo predictions of Table 1, extrapolated onto SLAC case of $R$=15 cm and variable energy.

Another example is the set of simulations made at 855 MeV with another Monte Carlo code [58]. Mazzolari et al produced for each bending radius of Si (111) crystal two predictions for electron dechanneling length at 855 MeV. One prediction was for the length observable in the experiment, where an electron was allowed in the simulations to exit the channeled state and possibly re-enter the channeled state again (so-called rechanneling).

Another prediction was for the so-called dynamic dechanneling length, which is the length until the first time the electron is dechanneled. This length is not possible to observe in the experiment, but it can be obtained in simulations. In Tables 1 and 2 and in Figure 1 we show both predictions. The dynamic dechanneling length is marked in our Tables as "$L_D$ dyn.".

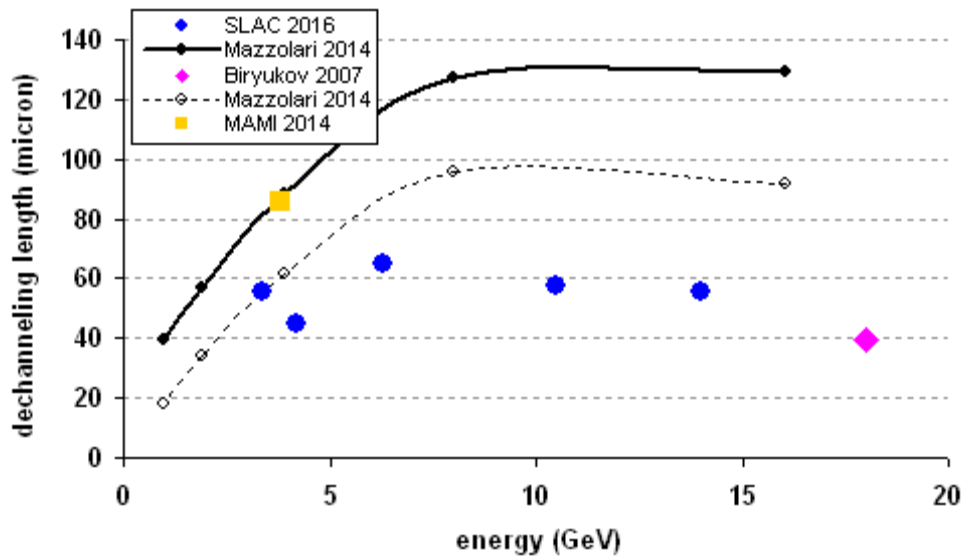

**Figure 1** SLAC data [59] on $L_D(E)$ is shown together with extrapolation of $L_D$ from MAMI experiment [58] and from two simulations, Mazzolari et al (solid line for $L_D$ and dashed line for dynamic $L_D$) [58] and Biryukov [62].

The MAMI experiment has measured the electron dechanneling length $L_D$ = 19.2 μm at 855 MeV in the Si (111) crystal planes bent with radius 3.35 cm. Notice that the 855-MeV simulations were made simultaneously with the experiment by the same authors. The agreement between what they measured (19.2 μm) and what they simulated (19.5 μm) for the dechanneling length is perfect.

Using Eq. (2), we extrapolated the 855 MeV predictions and measurements onto the case of SLAC crystal with $R$=15 cm in the same way as we earlier did with 50-GeV Monte Carlo prediction to bring it into comparison with SLAC data. A constant energy of MAMI $E_{fix} = E_1 = 855$ MeV and variable radius $R_{var} = R_1$ in the 855-MeV simulations were transformed into the constant radius $R_{fix} = R_2 = 15$ cm and variable energy $E_{var} = E_2$, according to $E_{var} = R_{var} \cdot (E_{fix}/R_{fix})$, in the range studied at SLAC, keeping the same ratios $E/R$. The extrapolated values are shown in Table 2 and in Figure 1. The $L_D$ data obtained at MAMI (marked $L_{D,MAMI}(E_1,R_1)$) with constant energy 855 MeV and variable radius were transformed into the $L_D$ values corresponding to the SLAC case (marked $L_{D,SLAC}(E_2,R_2)$) at variable energy and constant radius $R$=15 cm according to the formula

$$L_{D,SLAC}(E_2,R_2) = L_{D,MAMI}(E_1,R_1) \cdot (E_2/E_1) = L_{D,MAMI}(E_1,R_1) \cdot (R_2/R_1) \tag{9}$$

From Figure 1 and Table 2 we see that the data points from 3-14 GeV experiment and from 50 GeV simulation are in agreement. The data points from 855 MeV studies are a bit higher than SLAC data. Overall agreement of the sub-GeV data with multi-GeV data is within a factor of 1.5-2 or so. The procedure that we used to compare the predictions to the measurements did not have any fitting parameters. Moreover, the Monte Carlo predictions were published before the experiments at SLAC were undertaken.

## 4. Logarithmic correction of dechanneling length

We believe that it is helpful to apply here a logarithmic correction to the linear dependence of electron dechanneling length $L_D$ on energy that we used so far. The easiest way to do it is by using the modern formula for the *rms* angle of multiple scattering $\theta_{ms}$, provided by Particle Data Group [66], instead of the older simplified expression (2) used by Baier et al. Instead of Eq. (2), we take

$$\theta_{ms} = E_s/E \cdot (L/X)^{0.5} \cdot (1+0.038 \cdot \ln(L/X)) \tag{10}$$

We see that formally we can reduce the new equation to the previous form if we introduce an "effective" radiation length $X_{eff}$.

$$X_{eff} = X/(1+0.038 \cdot \ln(L/X))^2 \tag{11}$$

Replacing $X$ with $X_{eff}$ in Eq.(4), we obtain a logarithmic factor which depends on the length where dechanneling process takes place.

$$L_D \approx (2U_0 E/E_S^2) \cdot X_{eff} \tag{12}$$

or

$$L_D \approx (2U_0 E/(E_s(1+0.038 \cdot \ln(L/X)))^2) \cdot X_{eff}$$

Here, one could take $L$ simply as a crystal sample size along the beam as it naturally scales in the experiments for physics reasons. Typically, with energy $E$ growth, the crystal longitudinal size $L$ grows approximately linearly with $E$. In the following, for our estimates we take the electron dechanneling length $L_D$ from Eq.(4) as a characteristic length to be used in this equation. We introduce a logarithmic correction factor $k$

$$k(L_1,L_2) = (1+0.038 \cdot \ln(L_1/X))/(1+0.038 \cdot \ln(L_2/X)) \qquad (13)$$

to be added to Eq.(8) when we extrapolate $L_D$ from one energy case to another. Now we calculate:

$$L_D(E_2,R_2) = L_D(E_1,R_1) \cdot (E_2/E_1) \cdot k(L_1,L_2) \qquad (14)$$

|  |  | $E$ (GeV) | $R$ (cm) | $E/R$ (GeV/cm) | $L_D^{\log}$ (μm) | $L_D^{\log}$ dyn. (μm) |
|---|---|---|---|---|---|---|
| Biryukov | 2007 | 18 | 15 | 1.2 | 44.1 | - |
| MAMI | 2014 | 3.83 | 15 | 0.26 | 73.1 | - |
| Mazzolari et al. | 2014 | 16.03 | 15 | 1.07 | 93.8 | 66.2 |
| Mazzolari et al. | 2014 | 8.02 | 15 | 0.53 | 100.2 | 74.9 |
| Mazzolari et al. | 2014 | 3.89 | 15 | 0.26 | 75.3 | 52.4 |
| Mazzolari et al. | 2014 | 1.91 | 15 | 0.13 | 52.3 | 31.5 |
| Mazzolari et al. | 2014 | 0.96 | 15 | 0.06 | 39.3 | 17.9 |

**Table 3** Datasets of Table 2 with logarithmic correction applied on $L_D$ predictions (marked in Table 3 as $L_D^{\log}$).

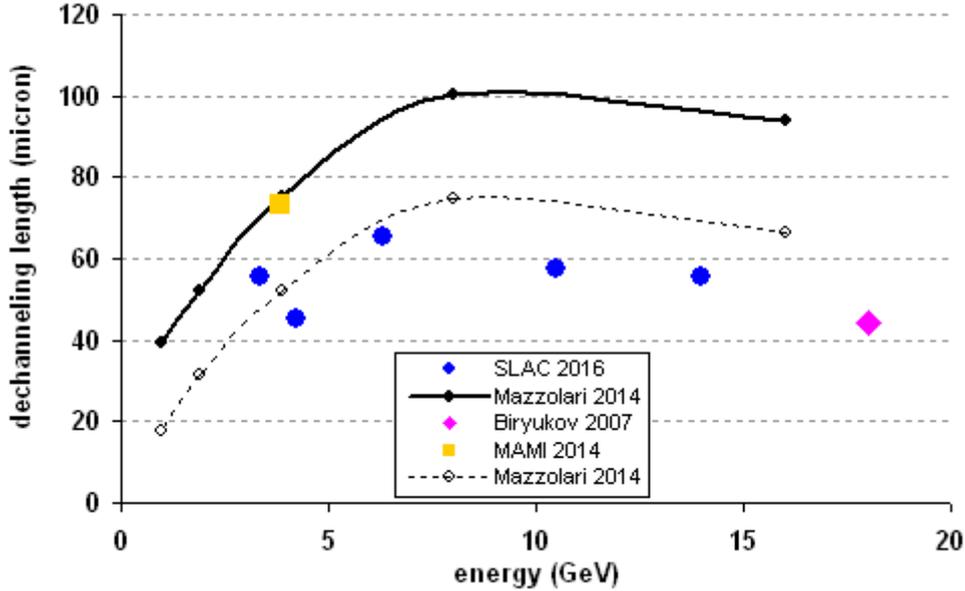

**Figure 2** The same as in Fig. 1 but the $L_D$ extrapolations are with logarithmic correction.

Table 3 shows the electron dechanneling length $L_D$ after the logarithmic correction applied to the 855-MeV and 50-GeV data to bring them into the range of SLAC. The corrected data are presented in Figure 2. This Figure 2 shows that the data from 855 MeV studies are now in better agreement with SLAC data. The data from 50-GeV simulation, where we applied the same logarithmic correction, are in good agreement with SLAC data again.

# 4. Conversion of $L_D$ dependence on energy into $L_D$ dependence on curvature

Also notice that we can transform a dataset with variable energy and constant radius into a dataset with a constant energy and variable radius and vice versa. As an example, we extrapolate also the SLAC measurements from the energy range 3.35-14 GeV with constant radius 15 cm onto the energy of MAMI, 855 MeV, with variable radius corresponding to the SLAC dataset. Table 4 shows the pairs of $E$, $R$ for SLAC and their equivalents for MAMI. Following equation (9), we calculate the dechanneling length for MAMI as a function of crystal radius. The extrapolated SLAC data are presented in Figure 3 together with the simulations made earlier directly for MAMI. The Monte Carlo prediction at 50 GeV extrapolated onto MAMI case is also shown in Table 4 and Fig. 3.

|         |      | $E$ (GeV) | $R$ (cm) | $E/R$ (GeV/cm) | $L_D$ (μm) | $E$ (GeV) | $R$ (cm) | $L_D$ (μm) |
|---------|------|-----------|----------|----------------|------------|-----------|----------|------------|
| SLAC    | 2016 | 3.35      | 15       | 0.22           | 55.4       | 0.855     | 3.83     | 16.5       |
| SLAC    | 2016 | 4.2       | 15       | 0.28           | 45.2       | 0.855     | 3.05     | 10.5       |
| SLAC    | 2016 | 6.3       | 15       | 0.42           | 65.3       | 0.855     | 2.04     | 10.5       |
| SLAC    | 2016 | 10.5      | 15       | 0.7            | 57.5       | 0.855     | 1.22     | 5.5        |
| SLAC    | 2016 | 14        | 15       | 0.93           | 55.8       | 0.855     | 0.92     | 4.0        |
| Biryukov| 2007 | 50        | 41.67    | 1.2            | 110        | 0.855     | 0.71     | 2.4        |

**Table 4** The SLAC $L_D(E)$ data at constant $R$=15 cm (and MC prediction at 50 GeV) converted into $L_D(R)$ values at $E$=855 MeV and variable $R$.

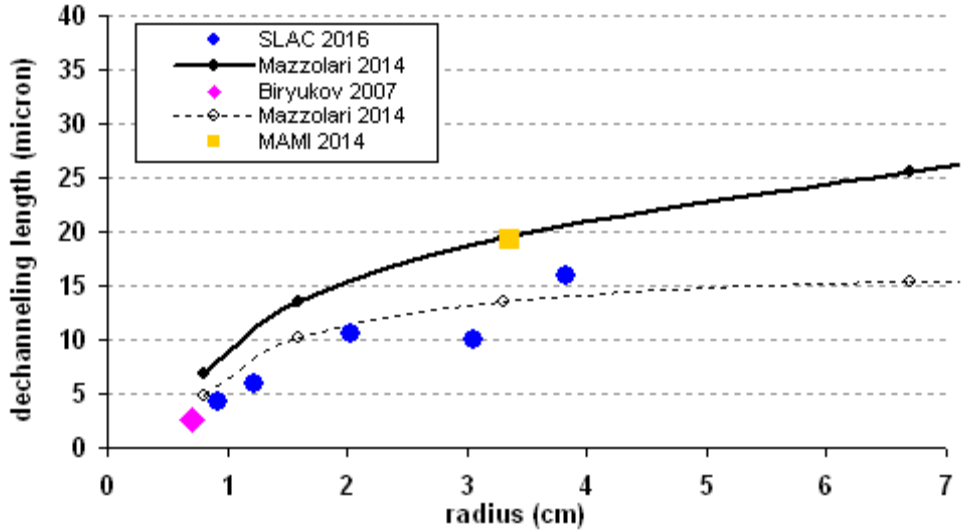

**Figure 3** The SLAC $L_D(E)$ data measured at $R$=15 cm (and MC prediction at 50 GeV) are extrapolated into $L_D(R)$ values at $E$=855 MeV and variable $R$. Compared with MAMI experiment and simulations of Mazzolari et al. at 855 MeV.

## 5. Conclusion

To conclude, there is a reasonable agreement between the predictions of Monte Carlo codes and the experiments on electron dechanneling length in bent crystals in sub-GeV and multi-GeV range. The data obtained in different experiments and Monte Carlo simulations can be cautiously extrapolated onto different ranges of energy and bending radius of the crystal. If the energy range exceeds an order of magnitude, a logarithmic correction to the electron dechanneling length seems necessary.